\documentclass[prl,twocolumn,superscriptaddress,longbibliography,preprintnumbers,floatfix,nofootinbib]{revtex4-1}

\usepackage{url}
\usepackage{xspace}
\usepackage{dsfont}
\usepackage{amssymb}
\usepackage{amsmath}
\usepackage{graphicx}
\usepackage[caption=false]{subfig}
\usepackage[colorlinks=true,citecolor=blue]{hyperref}
\usepackage{multirow}
\usepackage{hhline}

\usepackage{color}
\definecolor{mit-red}{rgb}{0.64,.12,0.2}
\definecolor{darkred}{rgb}{1.0,0.1,0.1}
\definecolor{darkgreen}{rgb}{0.1,0.7,0.1}
\definecolor{darkblue}{rgb}{0.1,0.1,1.0}

\usepackage{amsmath}

\newcommand{\sectionPRL}[1]{ \textbf{ #1.}}

\usepackage{tikz}
\usetikzlibrary{tikzmark}
\usetikzlibrary{calc} 
\usetikzlibrary{decorations.pathmorphing}
\usetikzlibrary{shapes.symbols}
\usetikzlibrary{backgrounds}
\usetikzlibrary{positioning}
\tikzstyle{box} = [rectangle, minimum width=0cm, minimum height=0.0cm, text centered, draw=black, fill=black!00, inner sep=1pt, inner ysep=1pt]

\begin{document}

\title{Morphing parton showers with event derivatives}

\author{Benjamin Nachman}
\email{bpnachman@lbl.gov}
\affiliation{Physics Division, Lawrence Berkeley National Laboratory, Berkeley, CA 94720, USA}
\affiliation{Berkeley Institute for Data Science, University of California, Berkeley, CA 94720, USA}

\author{Stefan Prestel}
\email{stefan.prestel.work@gmail.com}
\affiliation{Department of Astronomy and Theoretical Physics, Lund University, S-223 62 Lund, Sweden}

\begin{abstract}
We develop \texttt{EventMover}, a differentiable parton shower event generator.  This tool generates high- and variable-length scattering events that can be moved with simulation derivatives to change the value of the scale $\Lambda_\mathrm{QCD}$ defining the strong coupling constant, without introducing statistical variations between samples. To demonstrate the potential for \texttt{EventMover}, we compare the output of the simulation with $e^+e^-$ data to show how one could fit $\Lambda_\mathrm{QCD}$ with only a single event sample.  This is a critical step towards a fully differentiable event generator for particle and nuclear physics.
\end{abstract}

\maketitle

\sectionPRL{Introduction}
Simulations are essential tools for parameter estimation in particle and nuclear physics. Parton shower event generators furnish a model of the energy evolution of scattering processes, and are thus a central part of simulations. Currently, the parameter estimation proceeds in three steps.  First, a set of synthetic datasets are generated with various values of the parameters $\theta$.   Then, the experimental and synthetic data are passed through a dimension reduction step. Even though the data can be very high dimensional, typically each dataset is reduced to a one-dimensional representation (e.g., a histogram).  The dimensionality of the inference is limited because of the need to interpolate precisely between simulations produced with the coarsely spaced $\theta$ values.  Finally, the reduced representations of data and simulation are compared.  The $\theta$ corresponding to the synthetic dataset that is the best match to data is declared the fitted value.  Depending on the definition of `best match', synthetic datasets from nearby parameter values are then used to estimate uncertainties.

This paradigm significantly limits the potential to leverage data. The dimensional reduction of data often averages away important features, and may lead to suboptimal statistics to test the parameters of interest.  To make the most of complex particle and nuclear physics data, we need to use the full events in their natural high-dimensionality.  The key challenge is being able to interpolate event samples between values of simulation parameters.  One method is to fit the simulation at values with a smooth function in the parameters $\theta$.  This is called a \textit{surrogate model}.   Low-dimensional fits are common place in particle and nuclear physics. Deep learning methods are required to fit high-dimensional data with complex structure.  Given a differentiable surrogate model, one can perform gradient descent for optimization.  This approach has been explored for simulation/detector tuning~\cite{Andreassen:2019nnm,NEURIPS2020_a878dbeb} and effective field theory analysis~\cite{Brehmer:2018eca,Brehmer:2018kdj,Brehmer:2018hga,Brehmer:2019xox}.

The main drawbacks of surrogate modeling are that many simulation runs are necessary for high-fidelity (yet always approximate) fits. In some special cases, analytic tools exist to automatically morph a simulation with $\theta_0$ into a simulation with $\theta_1$, by means of ``reweighting".  For example, if the value of the strong coupling constant in a parton shower is varied, then event weights can be derived to adjust the relative event mixture~\cite{Mrenna:2016sih,Bellm:2016voq,Bothmann:2016nao}.  While highly useful, this does not change the events themselves, as individual simulations would have done. The statistical power of the dataset is diluted as the weights move away from unity.  Furthermore, there is no gradient information for relating events with nearby parameter values.  

A general approach that solves the challenges of surrogate modeling is to make the simulation itself differentiable.  A simulation is differentiable if it is efficient to compute derivatives with respect to the input parameters.  When we refer to differentiability, we are specifically referring to \textit{automatic differentiation} (autodiff) whereby derivatives are tracked through the simulation function and can readily achieve machine precision.  There are a variety of standard \texttt{C++} and \texttt{Python} tools for performing autodiff.  The \texttt{Python} tools are particularly widespread because they are naturally compatible with deep learning and running on Graphical Processing Units (GPUs).  These tools include \texttt{TensorFlow}~\cite{tensorflow2015-whitepaper}, \texttt{JAX}~\cite{jax2018github}, and \texttt{PyTorch}~\cite{NEURIPS2019_9015}.  The newest of these, \texttt{JAX}, is particularly popular because it provides an interface that is a drop-in replacement for \texttt{numpy}~\cite{harris2020array} functions and thus requires the least new syntax.

Our vision is for a fully differentiable event generator capable of comprehensively modeling scattering processes.  This letter represents a significant step towards this goal by introducing \texttt{EventMover}, the first differentiable parton shower.  For momentum transfer $Q^2\gg 1$ GeV, the phase space is mostly filled by final state radiation through showering.   Therefore, we can capture the complex high- and variable-dimensional nature of scattering events with \texttt{EventMover}.  The ultimate differentiable event generator will also incorporate hadronization and matrix element (ME) generation.  Hadronization cannot be modeled with first-principles simulations, making it natural to replace parameterised models directly with surrogates that can be tuned to data~\cite{Ilten:2022jfm,Ghosh:2022zdz}.  Differentiable MEs based on \texttt{MadGraph}~\cite{Alwall:2014hca} have been proposed in Ref.~\cite{Carrazza:2021gpx,Heinrich:2022xfa}.  An analogous differentiable simulation program is currently underway in cosmology~\cite{Modi:2020dyb,Bohm:2020ilt,Dai:2020ekz}.

\sectionPRL{Differentiable Simulation}
To illustrate how a simulation can be made differentiable, consider a Gaussian random variable $X\sim\mathcal{N}(\mu,\sigma)$ for mean $\mu$ and standard deviation $\sigma$.
 We can make this simulation differentiable by separating the randomness from the model parameters.  Let $Z$ be a uniform random variable between 0 and 1.  Then, $\phi(Z,\mu,\sigma)=\sigma \Phi^{-1}(Z)+\mu$ will have the same probability density as $X$, where $\Phi$ is the Gaussian Cumulative Distribution Function (CDF).  Writing the simulator this way has the feature that the random variables $Z$ do not depend on the model parameters $\mu$ and $\sigma$.  The simulator $z\rightarrow x=\phi(z,\mu,\sigma)$ is differentiable because we can compute $\partial x/\partial \mu$ and $\partial x/\partial \sigma$.  

\begin{figure}[h!]
\centering
\includegraphics[width=0.45\textwidth]{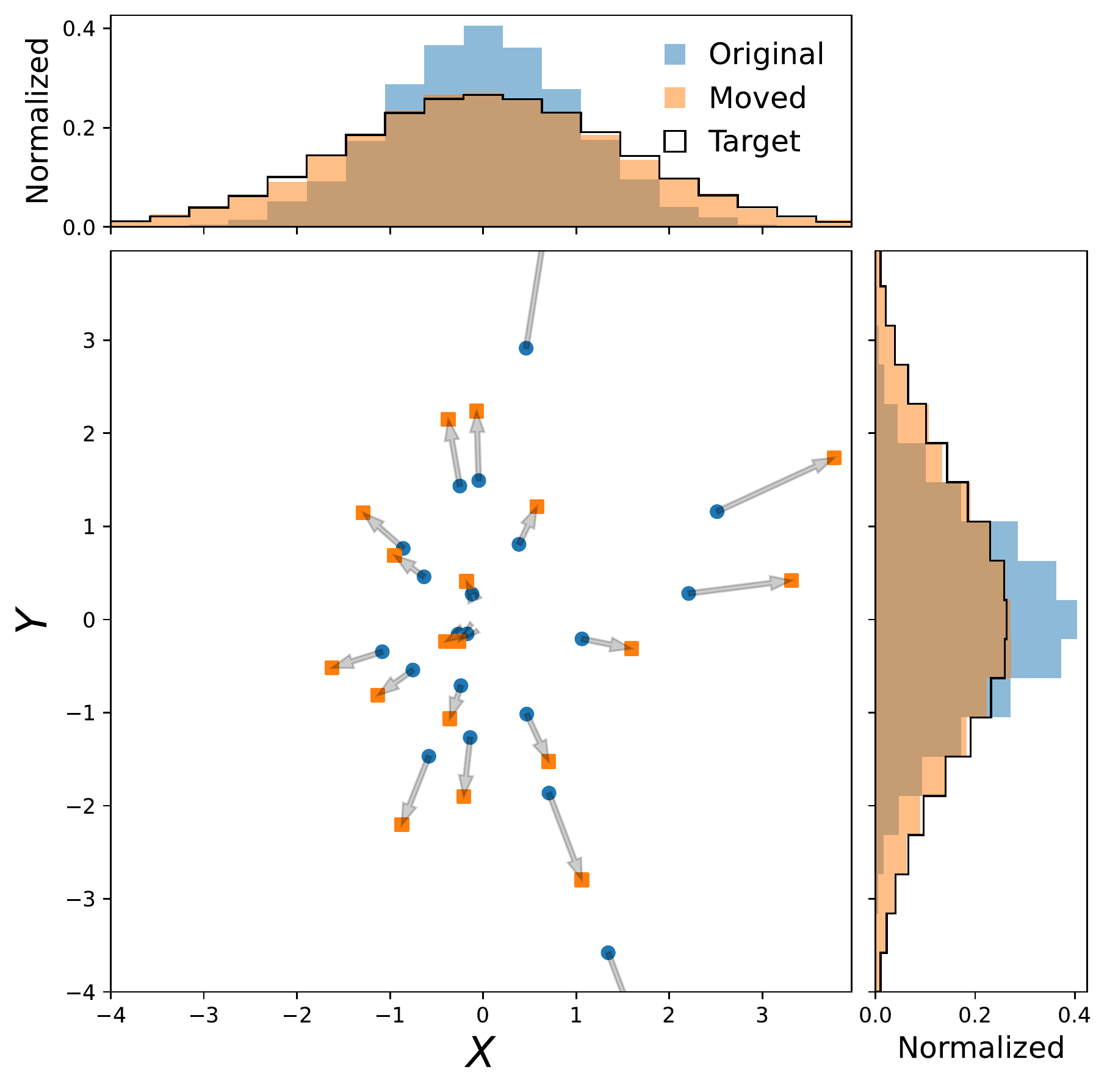}
\vskip -4mm
\caption{
\label{fig:gaussian}
An illustration of \texttt{EventMover} for a two-dimensional Gaussian simulation.  The original events are independent standard normal random variables, which are moved to independent Gaussians with zero mean $0$ and standard deviation $1.5$.  Projections of the two dimensions are shown at the top and on the right as histograms. The middle panel shows the trajectories of 20 events.}
\end{figure}

With the differentiable simulator $\phi$, we can smoothly move events.  For example, if we have an event sample $\{x_i\}$ generated with a particular value $(\mu_0,\sigma_0)$, we can simulateneously create a new event sample $\{x_i'\}$ with $x_i'=x_i+\nabla_\sigma\phi \,\Delta\sigma$ that will be statistically identical to a sample generated with $(\mu_0,\sigma_0+\Delta\sigma)$.
The gradient itself can also be computed efficiently (see \textit{back propagation}~\cite{Rumelhart:1986we}).  The moving of events in the Gaussian case is illustrated in Fig.~\ref{fig:gaussian} for $X\sim\mathcal{N}(\vec{0},I_2)\in\mathbb{R}^2$ where $I_2$ is the $2\times 2$ identity matrix and $\Delta\sigma=0.5$.  In this case, $\nabla_\sigma=[\Phi^{-1}(z_0),\Phi^{-1}(z_1)]$ so the further away a point starts from the origin, the more it gets moved.

\sectionPRL{Parton Shower Model}
Parton showers (PS) translate the ill-defined few-body scattering states into measurable asymptotic final states. They are crucial for any simulation of particle collisions (see e.g., Ref.~\cite{Buckley:2011ms,Hoche:2014rga}). The result of parton showering is variable particle-number scattering events $\vec e$ of a collection of particles, each of which is determined by an flavor, color, and four on-shell momentum quantum numbers. 

The parameters of the PS are correlated with the modeling of the highest-energy scattering as well as with the dynamics of hadronization. In conventional event generators, such correlations cannot easily be investigated, since parameter changes require individual simulations, which are subject to uncorrelated random noise. The underlying issue is algorithmic: PSs rely on an accept-reject method to sample states, thus requiring an undetermined quantity of random numbers. Parameter variations can change the random state of the system drastically. 

We have developed a new shower model that employs a fixed, well-defined quantity of random numbers. This model is based on the Discrete Quantum Chromodynamics (DQCD) method of~\cite{ANDERSSON1996217}, which has recently been employed in the context of quantum event generation~\cite{Gustafson:2022xwt}. We extend and improve this model to expose all latent variables and to enable automatic parameter variations through differentiable programming. The PS model depends on an overall mass scale $\Lambda$, which acts as parton-shower cut-off.  We set the reference value of the running coupling through the identification $\Lambda=\Lambda_\mathrm{QCD}$.

The DQCD method is based on the observation that gluons emitted from a color dipole act coherently if they are close enough in phase space.  To model this explicitly, the emission phase space is discretized.  In particular, the relative rapidity and relative transverse momentum are quantized.   Subsequent emissions introduce a fractal phase space (`grove', $g_\Lambda$) with each new piece shrinking until there is no room past $\Lambda$.  The kinematical properties in each discretized emission plane are used to compute the lab-frame momenta of the outgoing partons. As the number of possible emission histories is finite, there are a fixed number of possible random numbers needed to specify a state.   This is fundamentally different from conventional PSs, where the number of emissions is unbounded.  Going beyond previous implementations of DQCD, we set up the simulation code so that there is a one-to-one relation between random numbers and subsequent event generation, similarly to the Gaussian example from earlier. In essence, the coarse features of the result (number of emissions, phase-space regions assigned to the emissions) are selected \emph{before} the actual generation step.

The simple DQCD algorithm captures all features of soft gluon emission from color dipoles and is amenable to differentialization.  In particular, it is possible to compute gradients of both the event rate and the momenta of the outgoing partons with respect to $\Lambda$.  The gradient of the event rate is valid even if the shifted parameters lead to a changed phase space volume. The event rate depends very weakly on the parameters, so that the shifted rates are very narrowly peaked around the original weight. 

Crucially, we can now \emph{shift the kinematic properties of individual events} resulting from a change in $\Lambda$:
\begin{align}
\label{eq:event_expansion}
\vec e(g_\Lambda,\Lambda)\rightarrow\,\,& \vec e(g_{\Lambda'},\Lambda'=\Lambda+\Delta\Lambda)\\
&= \vec e(g_\Lambda,\Lambda) + \sum\limits_{n=1}^N \frac{\left(\Delta\Lambda\right)^n}{n!} \left. \frac{\partial^n \vec e}{\partial \Lambda^n}\right|_{\Lambda}\,,\nonumber
\end{align}
using autodiff to evaluate $\partial^n \vec e / \partial \Lambda^n$. These are the distinguishing features of \texttt{EventMover}. 

Differentiation with respect to the mass scale $\Lambda$ provide an excellent test of the algorithm, since $\Lambda$-variations change the phase space volume of the parton shower. Furthermore, information on the derivatives $\partial^n \vec e / \partial \Lambda^n$ allows to infer $\Lambda$ from experimental data, and thus define an extraction of the QCD coupling.

Changes in $\Lambda$ explicitly modify the phase space in the groves and implicitly modify the particle momenta via the two-particle invariant masses: $m_{ij}^2=\Lambda^2 \exp\left( \lambda_{ij}(g_\Lambda) \right)$, where $\lambda_{ij}(g_\Lambda)$ is the shortest distance between the two tips $i$ and $j$ along the grove graph. The $\Lambda$-dependence of $\lambda_{ij}(g_\Lambda)$ is weak.
Overall momentum conservation in $\partial^n \vec e / \partial \Lambda^n$ is guaranteed, but physical (on-shell, positive-energy) momenta of individual particles are not. This is expected, since off-shell momenta are required to morph a physical event to another physical event with new particle directions. Nevertheless, physical momenta of particles in moved events $\vec e(g_{\Lambda'},\Lambda')$ may demand the inclusion of higher-order terms in Eq.~\ref{eq:event_expansion}. The required expansion order depends on the original kinematics of $\vec e(g_{\Lambda},\Lambda)$ and the size and direction of the shift $\Delta\Lambda$. In rare cases, $n>3$ is required. The calculation of expansion terms would be impractical without autodiff.

\begin{figure}[h!]
\centering
\includegraphics[width=0.45\textwidth]{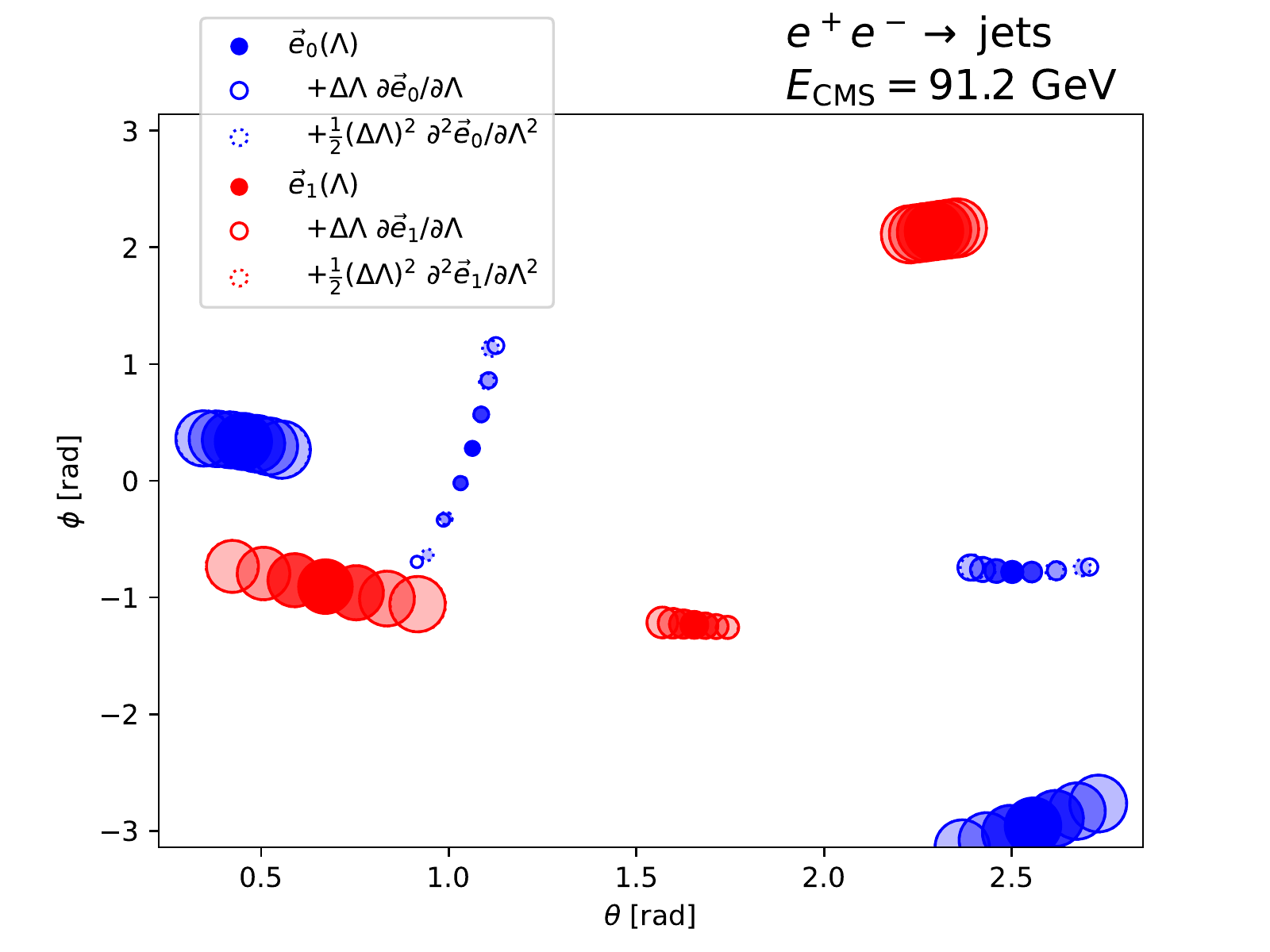}
\vskip -4mm
\caption{
\label{fig:event}
Event displays of two events. Each outgoing parton is represented by a circle with a radius proportional to its energy.  The darkest filled circles are the original particles and the lighter filled circles are the events simulated with different values of $\Lambda$ ($\pm 125,250,375$~MeV).  The unfilled circles indicate the events moved by the first derivative (solid line) and the second derivative (dotted line).  The circle outlines should match the filled circles.  This is true for nearly all particles except the low energy particles starting at $(\theta,\phi)\approx(1,0)$ and $(\theta,\phi)\approx(2.5,-1)$ where the second derivative terms are required for the most extreme shifts.  To help with visualization, all shifts are stretched by a factor of 5.}
\end{figure}

\sectionPRL{Results}
The main feature of \texttt{EventMover} is that it morphs events at one scale $\Lambda$ into events at another scale $\Lambda'$ using autodiff to realize Eq.~\ref{eq:event_expansion}. Examples of such ``moved" events are shown in Fig.~\ref{fig:event}. This shows that by employing Eq.~\ref{eq:event_expansion}, morphed events move smoothly and non-trivially across phase space. We find that for the bulk of events, first-order shifts are sufficient for most of the event with softer particles sometimes requiring second-order terms to produce physical moved events.

\begin{figure}[h!]
\includegraphics[width=0.4\textwidth]{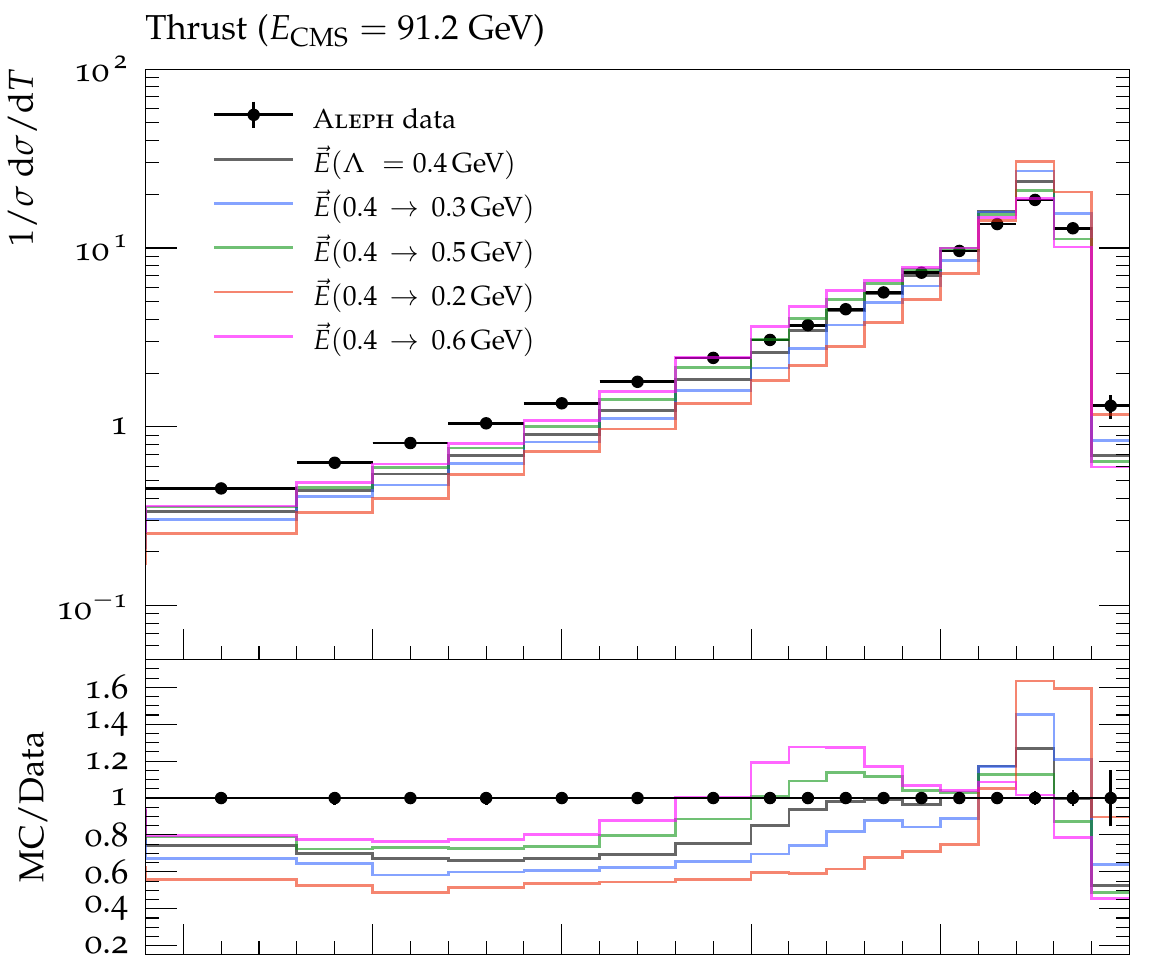}{}\\\vskip -1mm
\includegraphics[height=0.4\textwidth,angle=90]{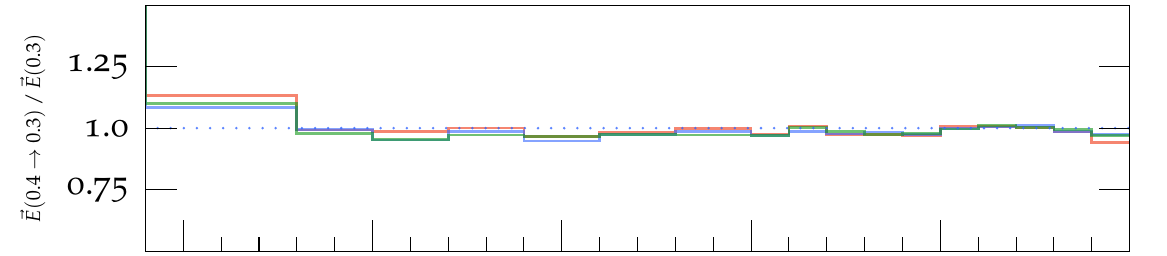}{}\\\vskip -1mm
\includegraphics[height=0.4\textwidth,angle=90]{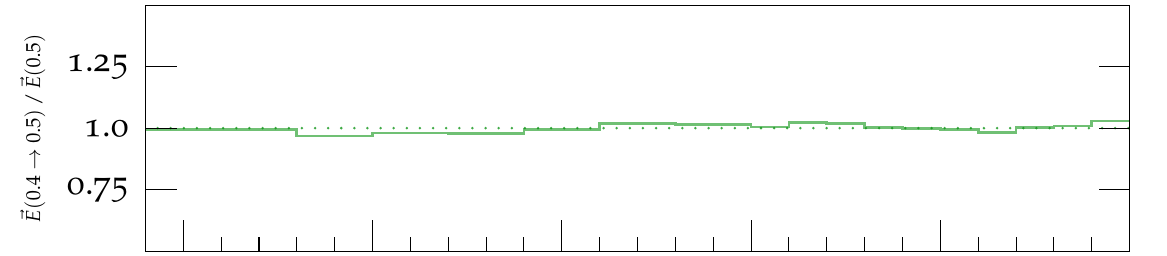}{}\\\vskip -1mm
\includegraphics[height=0.4\textwidth,angle=90]{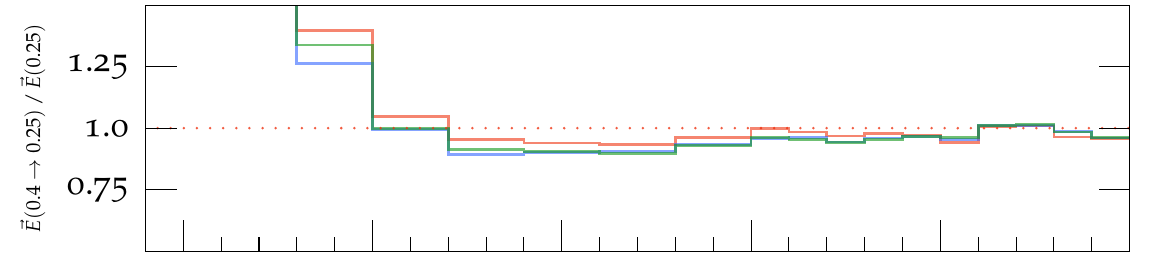}{}\\\vskip -1mm
\includegraphics[height=0.4\textwidth,angle=90]{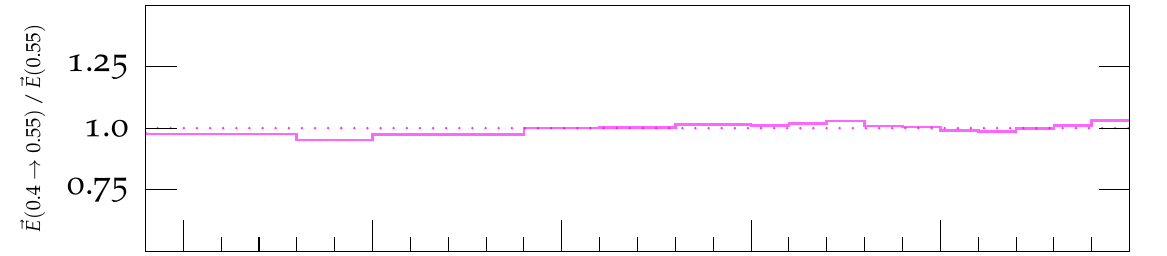}{}\\\vskip -1mm
\includegraphics[height=0.4\textwidth,angle=90]{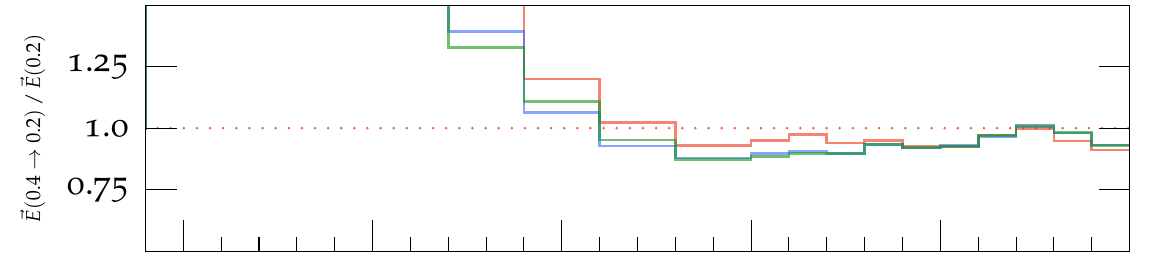}{}\\\vskip -1mm
\includegraphics[width=0.4\textwidth]{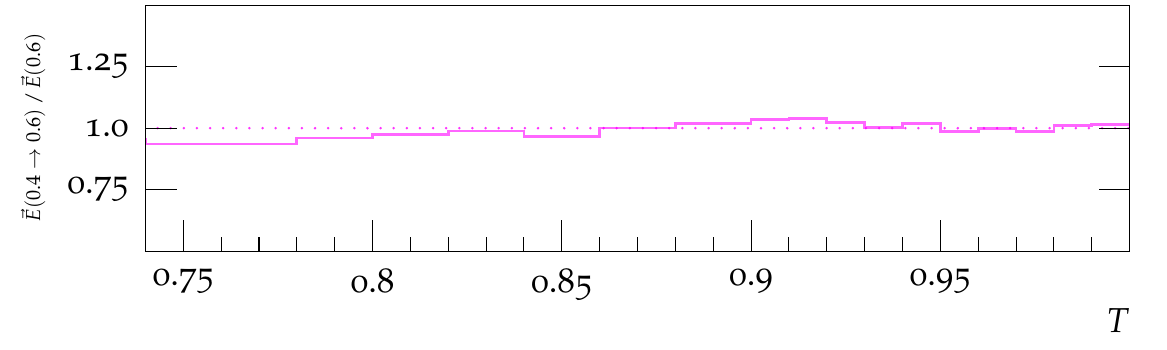}{}
\vskip -4mm
\caption{
\label{fig:thrust}
Sample results of \texttt{EventMover}, compared to \textsc{Aleph} data. The baseline event sample is indicated by $\vec E(\Lambda=0.4\,\mathrm{Gev})=\{\vec e_i(\Lambda) | i \in [1\dots N] \}$, while the $\vec E (0.4\rightarrow y \,\mathrm{Gev})$ labels events moved via autodiff. The second to sixth ratio show comparisons of moved and control event samples. In the second, fourth and sixth ratios, the red, blue and green curves show the effect of first-, second-, or third-order expansions.}
\end{figure}

The generation of full events allows for measurements of a plethora of observables that depend on final-state partons, jets, or hadrons.  For example, Fig.~\ref{fig:thrust} shows the spectrum of one of the most frequently studied event shape shape variables, thrust ($T$)~\cite{Farhi:1977sg}. Thrust-values of $T\sim 1$ (back-to-back jets) accounts for most of the cross section. The region $T\lesssim 3/4$ is highly sensitive to rare hard emissions and large invariant masses. The thrust variable provides an excellent laboratory to test the differentially moved events.  To compare with data, the events are passed through the string hadronization model~\cite{Andersson:1983ia} as implemented in \textsc{Pythia}~\cite{Bierlich:2022pfr}.

The baseline description of \texttt{EventMover} provides a satisfactory model of the \textsc{Aleph} data~\cite{ALEPH:2003obs} in Fig.~\ref{fig:thrust}. The most extreme event moves shift the baseline distribution by up to $\pm 50$\%, and are intentionally large. Statistical mismatches between the moved events and the control samples are vanishing. This would not be the case when reweighting, yet not moving, events~\cite{Mrenna:2016sih,Bellm:2016voq,Bothmann:2016nao}. 

For the bulk of events ($T\gtrsim 0.9$), moved events reproduce the respective control samples. For upward shifts of the mass scale, the control sample is reproduced throughout the whole spectrum. Downward shifts of the mass scale do, however, suffer from a pathological behavior. This observation is understood by dissecting the differentiation of the event kinematic properties.  Via the chain rule, the derivative of $\vec{e}$ with respect to $\Lambda$ is the gradient of the invariant-mass mapping $m_{ij}$ with respect to $\Lambda$ multiplied by the derivative of $\vec{e}$ with respect to the invariant masses.  The mass $m_{ij}^2(\Lambda)/\text{GeV}$ scales like $\Lambda^2/\text{GeV}$ which is much smaller than $\partial m_{ij}^2 /\partial\Lambda$, which scales like $2\Lambda$.  The invariants are then moved back to smaller values through the $m^2$-derivative of the phase space mapping. The latter converges reasonably fast for small shifts, including up-variations from a small $\Lambda$ value. For downwards shifts, the series converges slowly, since the moved events have to be ``dragged back" past the baseline value from very high intermediate values. In phase-space regions that are sensitive to large virtualities, the speed of convergence is slow, since the on-shell conditions get violated more severely by low-order terms. Overall, the expansion fails to converge for large downward shifts at high virtuality.  

\sectionPRL{Conclusions and Outlook}
In this letter, we have introduced \texttt{EventMover}, a final-state parton-shower algorithm. The algorithm is differential in its model parameters, meaning that each generated event may be accompanied by an arbitrarily large set of correlated moved events. Comparisons of \texttt{EventMover} to \textsc{Aleph} find reasonable agreement -- especially considering the approximations inherited from the Discrete QCD method.

We find that differentiable programming is crucial to produce high-fidelity moved events: moving events amounts to including sensitivity to derivatives of phase-space mappings, which lead to off-shell momenta -- as necessary to ``move" momentum directions. If this offshellness persists after first-order expansion, higher-order terms need to be included. The bulk of events can be described by including first- and second-order terms. We find that moving events to higher mass scales converges rapidly, while large downward shifts in extreme regions converge poorly.
Variations of the scale $\Lambda$ are of phenomenological interest, since $\Lambda$ is monotonically related to the strong coupling constant $\alpha_S$.
Thus, differentiation with respect to\ $\Lambda$ would allow for inference of $\alpha_S$ from data with a single simulated event sample.  Fits of this sort could even use machine learning models for goodness of fit statistics (see e.g.~Ref.~\cite{Andreassen:2019nnm}) or other differentiable statistical analysis methods~\cite{deCastro:2018mgh,Simpson:2022suz}.

The main reason to develop a differentiable parton shower is to allow for a straightforward inference of its parameters and uncertainties. \texttt{EventMover} builds on the Discrete QCD method, which employs a phase-space discretization derived from leading-logarithmic soft gluon resummation in the limit of infinite colors. Although some improvements of this method are conceivable (see e.g.~\cite{Gustafson:1992uh}), it is unlikely that higher-order QCD corrections could be derived analytically.  One direction for improvement could be to infer the `correct' phase space discretization (and rate) of higher-order corrections from comparison to analytic calculations. We believe \texttt{EventMover} is excellent candidate to build a precision parton shower on the Discrete-QCD paradigm, and hope that our method will inspire innovation enabling differentiable programming in traditional accept-reject-based showers on the quest for precision predictions~\cite{Campbell:2022qmc}.

\section*{Code and Data}

The \texttt{EventMover} code is available from \href{https://gitlab.com/discreteqcd/eventmover}{gitlab.com/discreteqcd/eventmover}. By virtue of \texttt{JAX}, the code may be run on CPUs and GPUs.

\section*{Acknowledgments}

BN is supported by the U.S. Department of Energy (DOE), Office of Science under contract DE-AC02-05CH11231. SP would like to thank G.~Gustafson and S.~Williams for discussions about Discrete QCD.
\textbf{}
\bibliography{refs,HEPML}

\begin{thebibliography}{34}%
\makeatletter
\providecommand \@ifxundefined [1]{%
 \@ifx{#1\undefined}
}%
\providecommand \@ifnum [1]{%
 \ifnum #1\expandafter \@firstoftwo
 \else \expandafter \@secondoftwo
 \fi
}%
\providecommand \@ifx [1]{%
 \ifx #1\expandafter \@firstoftwo
 \else \expandafter \@secondoftwo
 \fi
}%
\providecommand \natexlab [1]{#1}%
\providecommand \enquote  [1]{``#1''}%
\providecommand \bibnamefont  [1]{#1}%
\providecommand \bibfnamefont [1]{#1}%
\providecommand \citenamefont [1]{#1}%
\providecommand \href@noop [0]{\@secondoftwo}%
\providecommand \href [0]{\begingroup \@sanitize@url \@href}%
\providecommand \@href[1]{\@@startlink{#1}\@@href}%
\providecommand \@@href[1]{\endgroup#1\@@endlink}%
\providecommand \@sanitize@url [0]{\catcode `\\12\catcode `\$12\catcode
  `\&12\catcode `\#12\catcode `\^12\catcode `\_12\catcode `\%12\relax}%
\providecommand \@@startlink[1]{}%
\providecommand \@@endlink[0]{}%
\providecommand \url  [0]{\begingroup\@sanitize@url \@url }%
\providecommand \@url [1]{\endgroup\@href {#1}{\urlprefix }}%
\providecommand \urlprefix  [0]{URL }%
\providecommand \Eprint [0]{\href }%
\providecommand \doibase [0]{http://dx.doi.org/}%
\providecommand \selectlanguage [0]{\@gobble}%
\providecommand \bibinfo  [0]{\@secondoftwo}%
\providecommand \bibfield  [0]{\@secondoftwo}%
\providecommand \translation [1]{[#1]}%
\providecommand \BibitemOpen [0]{}%
\providecommand \bibitemStop [0]{}%
\providecommand \bibitemNoStop [0]{.\EOS\space}%
\providecommand \EOS [0]{\spacefactor3000\relax}%
\providecommand \BibitemShut  [1]{\csname bibitem#1\endcsname}%
\let\auto@bib@innerbib\@empty
\bibitem [{\citenamefont {Andreassen}\ and\ \citenamefont
  {Nachman}(2020)}]{Andreassen:2019nnm}%
  \BibitemOpen
  \bibfield  {author} {\bibinfo {author} {\bibfnamefont {Anders}\ \bibnamefont
  {Andreassen}}\ and\ \bibinfo {author} {\bibfnamefont {Benjamin}\ \bibnamefont
  {Nachman}},\ }\bibfield  {title} {\enquote {\bibinfo {title} {{Neural
  Networks for Full Phase-space Reweighting and Parameter Tuning}},}\ }\href
  {\doibase 10.1103/PhysRevD.101.091901} {\bibfield  {journal} {\bibinfo
  {journal} {Phys. Rev. D}\ }\textbf {\bibinfo {volume} {101}},\ \bibinfo
  {pages} {091901} (\bibinfo {year} {2020})},\ \Eprint
  {http://arxiv.org/abs/1907.08209} {arXiv:1907.08209 [hep-ph]} \BibitemShut
  {NoStop}%
\bibitem [{\citenamefont {Shirobokov}\ \emph {et~al.}(2020)\citenamefont
  {Shirobokov}, \citenamefont {Belavin}, \citenamefont {Kagan}, \citenamefont
  {Ustyuzhanin},\ and\ \citenamefont {Baydin}}]{NEURIPS2020_a878dbeb}%
  \BibitemOpen
  \bibfield  {author} {\bibinfo {author} {\bibfnamefont {Sergey}\ \bibnamefont
  {Shirobokov}}, \bibinfo {author} {\bibfnamefont {Vladislav}\ \bibnamefont
  {Belavin}}, \bibinfo {author} {\bibfnamefont {Michael}\ \bibnamefont
  {Kagan}}, \bibinfo {author} {\bibfnamefont {Andrei}\ \bibnamefont
  {Ustyuzhanin}}, \ and\ \bibinfo {author} {\bibfnamefont {Atilim~Gunes}\
  \bibnamefont {Baydin}},\ }\bibfield  {title} {\enquote {\bibinfo {title}
  {{Black-Box Optimization with Local Generative Surrogates}},}\ }in\ \href
  {https://proceedings.neurips.cc/paper/2020/hash/a878dbebc902328b41dbf02aa87abb58-Abstract.html}
  {\emph {\bibinfo {booktitle} {{Advances in Neural Information Processing
  Systems}}}},\ Vol.~\bibinfo {volume} {33},\ \bibinfo {editor} {edited by\
  \bibinfo {editor} {\bibfnamefont {H.}~\bibnamefont {Larochelle}}, \bibinfo
  {editor} {\bibfnamefont {M.}~\bibnamefont {Ranzato}}, \bibinfo {editor}
  {\bibfnamefont {R.}~\bibnamefont {Hadsell}}, \bibinfo {editor} {\bibfnamefont
  {M.~F.}\ \bibnamefont {Balcan}}, \ and\ \bibinfo {editor} {\bibfnamefont
  {H.}~\bibnamefont {Lin}}}\ (\bibinfo  {publisher} {Curran Associates, Inc.},\
  \bibinfo {year} {2020})\ pp.\ \bibinfo {pages} {14650--14662},\ \Eprint
  {http://arxiv.org/abs/2002.04632} {arXiv:2002.04632 [cs.LG]} \BibitemShut
  {NoStop}%
\bibitem [{\citenamefont {Brehmer}\ \emph
  {et~al.}(2018{\natexlab{a}})\citenamefont {Brehmer}, \citenamefont {Cranmer},
  \citenamefont {Louppe},\ and\ \citenamefont {Pavez}}]{Brehmer:2018eca}%
  \BibitemOpen
  \bibfield  {author} {\bibinfo {author} {\bibfnamefont {Johann}\ \bibnamefont
  {Brehmer}}, \bibinfo {author} {\bibfnamefont {Kyle}\ \bibnamefont {Cranmer}},
  \bibinfo {author} {\bibfnamefont {Gilles}\ \bibnamefont {Louppe}}, \ and\
  \bibinfo {author} {\bibfnamefont {Juan}\ \bibnamefont {Pavez}},\ }\bibfield
  {title} {\enquote {\bibinfo {title} {{A Guide to Constraining Effective Field
  Theories with Machine Learning}},}\ }\href {\doibase
  10.1103/PhysRevD.98.052004} {\  (\bibinfo {year} {2018}{\natexlab{a}}),\
  10.1103/PhysRevD.98.052004},\ \Eprint {http://arxiv.org/abs/1805.00020}
  {arXiv:1805.00020 [hep-ph]} \BibitemShut {NoStop}%
\bibitem [{\citenamefont {Brehmer}\ \emph
  {et~al.}(2018{\natexlab{b}})\citenamefont {Brehmer}, \citenamefont {Cranmer},
  \citenamefont {Louppe},\ and\ \citenamefont {Pavez}}]{Brehmer:2018kdj}%
  \BibitemOpen
  \bibfield  {author} {\bibinfo {author} {\bibfnamefont {Johann}\ \bibnamefont
  {Brehmer}}, \bibinfo {author} {\bibfnamefont {Kyle}\ \bibnamefont {Cranmer}},
  \bibinfo {author} {\bibfnamefont {Gilles}\ \bibnamefont {Louppe}}, \ and\
  \bibinfo {author} {\bibfnamefont {Juan}\ \bibnamefont {Pavez}},\ }\bibfield
  {title} {\enquote {\bibinfo {title} {{Constraining Effective Field Theories
  with Machine Learning}},}\ }\href {\doibase 10.1103/PhysRevLett.121.111801}
  {\  (\bibinfo {year} {2018}{\natexlab{b}}),\
  10.1103/PhysRevLett.121.111801},\ \Eprint {http://arxiv.org/abs/1805.00013}
  {arXiv:1805.00013 [hep-ph]} \BibitemShut {NoStop}%
\bibitem [{\citenamefont {Brehmer}\ \emph
  {et~al.}(2020{\natexlab{a}})\citenamefont {Brehmer}, \citenamefont {Louppe},
  \citenamefont {Pavez},\ and\ \citenamefont {Cranmer}}]{Brehmer:2018hga}%
  \BibitemOpen
  \bibfield  {author} {\bibinfo {author} {\bibfnamefont {Johann}\ \bibnamefont
  {Brehmer}}, \bibinfo {author} {\bibfnamefont {Gilles}\ \bibnamefont
  {Louppe}}, \bibinfo {author} {\bibfnamefont {Juan}\ \bibnamefont {Pavez}}, \
  and\ \bibinfo {author} {\bibfnamefont {Kyle}\ \bibnamefont {Cranmer}},\
  }\bibfield  {title} {\enquote {\bibinfo {title} {{Mining gold from implicit
  models to improve likelihood-free inference}},}\ }\href {\doibase
  10.1073/pnas.1915980117} {\bibfield  {journal} {\bibinfo  {journal} {Proc.
  Nat. Acad. Sci.}\ ,\ \bibinfo {pages} {201915980}} (\bibinfo {year}
  {2020}{\natexlab{a}})},\ \Eprint {http://arxiv.org/abs/1805.12244}
  {arXiv:1805.12244 [stat.ML]} \BibitemShut {NoStop}%
\bibitem [{\citenamefont {Brehmer}\ \emph
  {et~al.}(2020{\natexlab{b}})\citenamefont {Brehmer}, \citenamefont {Kling},
  \citenamefont {Espejo},\ and\ \citenamefont {Cranmer}}]{Brehmer:2019xox}%
  \BibitemOpen
  \bibfield  {author} {\bibinfo {author} {\bibfnamefont {Johann}\ \bibnamefont
  {Brehmer}}, \bibinfo {author} {\bibfnamefont {Felix}\ \bibnamefont {Kling}},
  \bibinfo {author} {\bibfnamefont {Irina}\ \bibnamefont {Espejo}}, \ and\
  \bibinfo {author} {\bibfnamefont {Kyle}\ \bibnamefont {Cranmer}},\ }\bibfield
   {title} {\enquote {\bibinfo {title} {{MadMiner: Machine learning-based
  inference for particle physics}},}\ }\href {\doibase
  10.1007/s41781-020-0035-2} {\bibfield  {journal} {\bibinfo  {journal}
  {Comput. Softw. Big Sci.}\ }\textbf {\bibinfo {volume} {4}},\ \bibinfo
  {pages} {3} (\bibinfo {year} {2020}{\natexlab{b}})},\ \Eprint
  {http://arxiv.org/abs/1907.10621} {arXiv:1907.10621 [hep-ph]} \BibitemShut
  {NoStop}%
\bibitem [{\citenamefont {Mrenna}\ and\ \citenamefont
  {Skands}(2016)}]{Mrenna:2016sih}%
  \BibitemOpen
  \bibfield  {author} {\bibinfo {author} {\bibfnamefont {S.}~\bibnamefont
  {Mrenna}}\ and\ \bibinfo {author} {\bibfnamefont {P.}~\bibnamefont
  {Skands}},\ }\bibfield  {title} {\enquote {\bibinfo {title} {{Automated
  Parton-Shower Variations in Pythia 8}},}\ }\href {\doibase
  10.1103/PhysRevD.94.074005} {\bibfield  {journal} {\bibinfo  {journal} {Phys.
  Rev. D}\ }\textbf {\bibinfo {volume} {94}},\ \bibinfo {pages} {074005}
  (\bibinfo {year} {2016})},\ \Eprint {http://arxiv.org/abs/1605.08352}
  {arXiv:1605.08352 [hep-ph]} \BibitemShut {NoStop}%
\bibitem [{\citenamefont {Bellm}\ \emph {et~al.}(2016)\citenamefont {Bellm},
  \citenamefont {Pl\"atzer}, \citenamefont {Richardson}, \citenamefont
  {Si\'odmok},\ and\ \citenamefont {Webster}}]{Bellm:2016voq}%
  \BibitemOpen
  \bibfield  {author} {\bibinfo {author} {\bibfnamefont {Johannes}\
  \bibnamefont {Bellm}}, \bibinfo {author} {\bibfnamefont {Simon}\ \bibnamefont
  {Pl\"atzer}}, \bibinfo {author} {\bibfnamefont {Peter}\ \bibnamefont
  {Richardson}}, \bibinfo {author} {\bibfnamefont {Andrzej}\ \bibnamefont
  {Si\'odmok}}, \ and\ \bibinfo {author} {\bibfnamefont {Stephen}\ \bibnamefont
  {Webster}},\ }\bibfield  {title} {\enquote {\bibinfo {title} {{Reweighting
  Parton Showers}},}\ }\href {\doibase 10.1103/PhysRevD.94.034028} {\bibfield
  {journal} {\bibinfo  {journal} {Phys. Rev. D}\ }\textbf {\bibinfo {volume}
  {94}},\ \bibinfo {pages} {034028} (\bibinfo {year} {2016})},\ \Eprint
  {http://arxiv.org/abs/1605.08256} {arXiv:1605.08256 [hep-ph]} \BibitemShut
  {NoStop}%
\bibitem [{\citenamefont {Bothmann}\ \emph {et~al.}(2016)\citenamefont
  {Bothmann}, \citenamefont {Sch\"onherr},\ and\ \citenamefont
  {Schumann}}]{Bothmann:2016nao}%
  \BibitemOpen
  \bibfield  {author} {\bibinfo {author} {\bibfnamefont {Enrico}\ \bibnamefont
  {Bothmann}}, \bibinfo {author} {\bibfnamefont {Marek}\ \bibnamefont
  {Sch\"onherr}}, \ and\ \bibinfo {author} {\bibfnamefont {Steffen}\
  \bibnamefont {Schumann}},\ }\bibfield  {title} {\enquote {\bibinfo {title}
  {{Reweighting QCD matrix-element and parton-shower calculations}},}\ }\href
  {\doibase 10.1140/epjc/s10052-016-4430-0} {\bibfield  {journal} {\bibinfo
  {journal} {Eur. Phys. J. C}\ }\textbf {\bibinfo {volume} {76}},\ \bibinfo
  {pages} {590} (\bibinfo {year} {2016})},\ \Eprint
  {http://arxiv.org/abs/1606.08753} {arXiv:1606.08753 [hep-ph]} \BibitemShut
  {NoStop}%
\bibitem [{\citenamefont {Abadi}\ \emph {et~al.}(2015)\citenamefont {Abadi}
  \emph {et~al.}}]{tensorflow2015-whitepaper}%
  \BibitemOpen
  \bibfield  {author} {\bibinfo {author} {\bibfnamefont {Mart\'{\i}n}\
  \bibnamefont {Abadi}} \emph {et~al.},\ }\href {http://tensorflow.org/}
  {\enquote {\bibinfo {title} {{TensorFlow}: Large-scale machine learning on
  heterogeneous systems},}\ } (\bibinfo {year} {2015}),\ \bibinfo {note}
  {software available from tensorflow.org}\BibitemShut {NoStop}%
\bibitem [{\citenamefont {Bradbury}\ \emph {et~al.}(2018)\citenamefont
  {Bradbury} \emph {et~al.}}]{jax2018github}%
  \BibitemOpen
  \bibfield  {author} {\bibinfo {author} {\bibfnamefont {James}\ \bibnamefont
  {Bradbury}} \emph {et~al.},\ }\href {http://github.com/google/jax} {\enquote
  {\bibinfo {title} {{JAX}: composable transformations of {P}ython+{N}um{P}y
  programs},}\ } (\bibinfo {year} {2018})\BibitemShut {NoStop}%
\bibitem [{\citenamefont {Paszke}\ \emph {et~al.}(2019)\citenamefont {Paszke}
  \emph {et~al.}}]{NEURIPS2019_9015}%
  \BibitemOpen
  \bibfield  {author} {\bibinfo {author} {\bibfnamefont {Adam}\ \bibnamefont
  {Paszke}} \emph {et~al.},\ }\bibfield  {title} {\enquote {\bibinfo {title}
  {Pytorch: An imperative style, high-performance deep learning library},}\
  }in\ \href
  {http://papers.neurips.cc/paper/9015-pytorch-an-imperative-style-high-performance-deep-learning-library.pdf}
  {\emph {\bibinfo {booktitle} {Advances in Neural Information Processing
  Systems 32}}},\ \bibinfo {editor} {edited by\ \bibinfo {editor}
  {\bibfnamefont {H.}~\bibnamefont {Wallach}}, \bibinfo {editor} {\bibfnamefont
  {H.}~\bibnamefont {Larochelle}}, \bibinfo {editor} {\bibfnamefont
  {A.}~\bibnamefont {Beygelzimer}}, \bibinfo {editor} {\bibfnamefont
  {F.}~\bibnamefont {d'Alch\'{e} Buc}}, \bibinfo {editor} {\bibfnamefont
  {E.}~\bibnamefont {Fox}}, \ and\ \bibinfo {editor} {\bibfnamefont
  {R.}~\bibnamefont {Garnett}}}\ (\bibinfo  {publisher} {Curran Associates,
  Inc.},\ \bibinfo {year} {2019})\ pp.\ \bibinfo {pages}
  {8024--8035}\BibitemShut {NoStop}%
\bibitem [{\citenamefont {Harris}\ \emph {et~al.}(2020)\citenamefont {Harris}
  \emph {et~al.}}]{harris2020array}%
  \BibitemOpen
  \bibfield  {author} {\bibinfo {author} {\bibfnamefont {Charles~R.}\
  \bibnamefont {Harris}} \emph {et~al.},\ }\bibfield  {title} {\enquote
  {\bibinfo {title} {Array programming with {NumPy}},}\ }\href {\doibase
  10.1038/s41586-020-2649-2} {\bibfield  {journal} {\bibinfo  {journal}
  {Nature}\ }\textbf {\bibinfo {volume} {585}},\ \bibinfo {pages} {357--362}
  (\bibinfo {year} {2020})}\BibitemShut {NoStop}%
\bibitem [{\citenamefont {Ilten}\ \emph {et~al.}(2022)\citenamefont {Ilten},
  \citenamefont {Menzo}, \citenamefont {Youssef},\ and\ \citenamefont
  {Zupan}}]{Ilten:2022jfm}%
  \BibitemOpen
  \bibfield  {author} {\bibinfo {author} {\bibfnamefont {Phil}\ \bibnamefont
  {Ilten}}, \bibinfo {author} {\bibfnamefont {Tony}\ \bibnamefont {Menzo}},
  \bibinfo {author} {\bibfnamefont {Ahmed}\ \bibnamefont {Youssef}}, \ and\
  \bibinfo {author} {\bibfnamefont {Jure}\ \bibnamefont {Zupan}},\ }\bibfield
  {title} {\enquote {\bibinfo {title} {{Modeling hadronization using machine
  learning}},}\ }\href@noop {} {\  (\bibinfo {year} {2022})},\ \Eprint
  {http://arxiv.org/abs/2203.04983} {arXiv:2203.04983 [hep-ph]} \BibitemShut
  {NoStop}%
\bibitem [{\citenamefont {Ghosh}\ \emph {et~al.}(2022)\citenamefont {Ghosh},
  \citenamefont {Ju}, \citenamefont {Nachman},\ and\ \citenamefont
  {Siodmok}}]{Ghosh:2022zdz}%
  \BibitemOpen
  \bibfield  {author} {\bibinfo {author} {\bibfnamefont {Aishik}\ \bibnamefont
  {Ghosh}}, \bibinfo {author} {\bibfnamefont {Xiangyang}\ \bibnamefont {Ju}},
  \bibinfo {author} {\bibfnamefont {Benjamin}\ \bibnamefont {Nachman}}, \ and\
  \bibinfo {author} {\bibfnamefont {Andrzej}\ \bibnamefont {Siodmok}},\
  }\bibfield  {title} {\enquote {\bibinfo {title} {{Towards a Deep Learning
  Model for Hadronization}},}\ }\href@noop {} {\  (\bibinfo {year} {2022})},\
  \Eprint {http://arxiv.org/abs/2203.12660} {arXiv:2203.12660 [hep-ph]}
  \BibitemShut {NoStop}%
\bibitem [{\citenamefont {Alwall}\ \emph {et~al.}(2014)\citenamefont {Alwall},
  \citenamefont {Frederix}, \citenamefont {Frixione}, \citenamefont {Hirschi},
  \citenamefont {Maltoni}, \citenamefont {Mattelaer}, \citenamefont {Shao},
  \citenamefont {Stelzer}, \citenamefont {Torrielli},\ and\ \citenamefont
  {Zaro}}]{Alwall:2014hca}%
  \BibitemOpen
  \bibfield  {author} {\bibinfo {author} {\bibfnamefont {J.}~\bibnamefont
  {Alwall}}, \bibinfo {author} {\bibfnamefont {R.}~\bibnamefont {Frederix}},
  \bibinfo {author} {\bibfnamefont {S.}~\bibnamefont {Frixione}}, \bibinfo
  {author} {\bibfnamefont {V.}~\bibnamefont {Hirschi}}, \bibinfo {author}
  {\bibfnamefont {F.}~\bibnamefont {Maltoni}}, \bibinfo {author} {\bibfnamefont
  {O.}~\bibnamefont {Mattelaer}}, \bibinfo {author} {\bibfnamefont {H.~S.}\
  \bibnamefont {Shao}}, \bibinfo {author} {\bibfnamefont {T.}~\bibnamefont
  {Stelzer}}, \bibinfo {author} {\bibfnamefont {P.}~\bibnamefont {Torrielli}},
  \ and\ \bibinfo {author} {\bibfnamefont {M.}~\bibnamefont {Zaro}},\
  }\bibfield  {title} {\enquote {\bibinfo {title} {{The automated computation
  of tree-level and next-to-leading order differential cross sections, and
  their matching to parton shower simulations}},}\ }\href {\doibase
  10.1007/JHEP07(2014)079} {\bibfield  {journal} {\bibinfo  {journal} {JHEP}\
  }\textbf {\bibinfo {volume} {07}},\ \bibinfo {pages} {079} (\bibinfo {year}
  {2014})},\ \Eprint {http://arxiv.org/abs/1405.0301} {arXiv:1405.0301
  [hep-ph]} \BibitemShut {NoStop}%
\bibitem [{\citenamefont {Carrazza}\ \emph {et~al.}(2021)\citenamefont
  {Carrazza}, \citenamefont {Cruz-Martinez}, \citenamefont {Rossi},\ and\
  \citenamefont {Zaro}}]{Carrazza:2021gpx}%
  \BibitemOpen
  \bibfield  {author} {\bibinfo {author} {\bibfnamefont {Stefano}\ \bibnamefont
  {Carrazza}}, \bibinfo {author} {\bibfnamefont {Juan}\ \bibnamefont
  {Cruz-Martinez}}, \bibinfo {author} {\bibfnamefont {Marco}\ \bibnamefont
  {Rossi}}, \ and\ \bibinfo {author} {\bibfnamefont {Marco}\ \bibnamefont
  {Zaro}},\ }\bibfield  {title} {\enquote {\bibinfo {title} {{MadFlow:
  automating Monte Carlo simulation on GPU for particle physics processes}},}\
  }\href {\doibase 10.1140/epjc/s10052-021-09443-8} {\bibfield  {journal}
  {\bibinfo  {journal} {Eur. Phys. J. C}\ }\textbf {\bibinfo {volume} {81}},\
  \bibinfo {pages} {656} (\bibinfo {year} {2021})},\ \Eprint
  {http://arxiv.org/abs/2106.10279} {arXiv:2106.10279 [physics.comp-ph]}
  \BibitemShut {NoStop}%
\bibitem [{\citenamefont {Heinrich}\ and\ \citenamefont
  {Kagan}(2022)}]{Heinrich:2022xfa}%
  \BibitemOpen
  \bibfield  {author} {\bibinfo {author} {\bibfnamefont {Lukas}\ \bibnamefont
  {Heinrich}}\ and\ \bibinfo {author} {\bibfnamefont {Michael}\ \bibnamefont
  {Kagan}},\ }\bibfield  {title} {\enquote {\bibinfo {title} {{Differentiable
  Matrix Elements with $MadJax$}},}\ }in\ \href@noop {} {\emph {\bibinfo
  {booktitle} {{20th International Workshop on Advanced Computing and Analysis
  Techniques in Physics Research}: {AI Decoded - Towards Sustainable, Diverse,
  Performant and Effective Scientific Computing}}}}\ (\bibinfo {year} {2022})\
  \Eprint {http://arxiv.org/abs/2203.00057} {arXiv:2203.00057 [hep-ph]}
  \BibitemShut {NoStop}%
\bibitem [{\citenamefont {Modi}\ \emph {et~al.}(2021)\citenamefont {Modi},
  \citenamefont {Lanusse},\ and\ \citenamefont {Seljak}}]{Modi:2020dyb}%
  \BibitemOpen
  \bibfield  {author} {\bibinfo {author} {\bibfnamefont {Chirag}\ \bibnamefont
  {Modi}}, \bibinfo {author} {\bibfnamefont {Francois}\ \bibnamefont
  {Lanusse}}, \ and\ \bibinfo {author} {\bibfnamefont {Uros}\ \bibnamefont
  {Seljak}},\ }\bibfield  {title} {\enquote {\bibinfo {title} {{FlowPM:
  Distributed TensorFlow implementation of the FastPM cosmological N-body
  solver}},}\ }\href {\doibase 10.1016/j.ascom.2021.100505} {\bibfield
  {journal} {\bibinfo  {journal} {Astron. Comput.}\ }\textbf {\bibinfo {volume}
  {37}},\ \bibinfo {pages} {100505} (\bibinfo {year} {2021})},\ \Eprint
  {http://arxiv.org/abs/2010.11847} {arXiv:2010.11847 [astro-ph.CO]}
  \BibitemShut {NoStop}%
\bibitem [{\citenamefont {B\"ohm}\ \emph {et~al.}(2021)\citenamefont {B\"ohm},
  \citenamefont {Feng}, \citenamefont {Lee},\ and\ \citenamefont
  {Dai}}]{Bohm:2020ilt}%
  \BibitemOpen
  \bibfield  {author} {\bibinfo {author} {\bibfnamefont {Vanessa}\ \bibnamefont
  {B\"ohm}}, \bibinfo {author} {\bibfnamefont {Yu}~\bibnamefont {Feng}},
  \bibinfo {author} {\bibfnamefont {Max~E.}\ \bibnamefont {Lee}}, \ and\
  \bibinfo {author} {\bibfnamefont {Biwei}\ \bibnamefont {Dai}},\ }\bibfield
  {title} {\enquote {\bibinfo {title} {{MADLens, a python package for fast and
  differentiable non-Gaussian lensing simulations}},}\ }\href {\doibase
  10.1016/j.ascom.2021.100490} {\bibfield  {journal} {\bibinfo  {journal}
  {Astron. Comput.}\ }\textbf {\bibinfo {volume} {36}},\ \bibinfo {pages}
  {100490} (\bibinfo {year} {2021})},\ \Eprint
  {http://arxiv.org/abs/2012.07266} {arXiv:2012.07266 [astro-ph.CO]}
  \BibitemShut {NoStop}%
\bibitem [{\citenamefont {Dai}\ and\ \citenamefont
  {Seljak}(2020)}]{Dai:2020ekz}%
  \BibitemOpen
  \bibfield  {author} {\bibinfo {author} {\bibfnamefont {Biwei}\ \bibnamefont
  {Dai}}\ and\ \bibinfo {author} {\bibfnamefont {Uro\v{s}}\ \bibnamefont
  {Seljak}},\ }\bibfield  {title} {\enquote {\bibinfo {title} {{Learning
  effective physical laws for generating cosmological hydrodynamics with
  Lagrangian Deep Learning}},}\ }\href {\doibase 10.1073/pnas.2020324118} {\
  (\bibinfo {year} {2020}),\ 10.1073/pnas.2020324118},\ \Eprint
  {http://arxiv.org/abs/2010.02926} {arXiv:2010.02926 [astro-ph.CO]}
  \BibitemShut {NoStop}%
\bibitem [{\citenamefont {Rumelhart}\ \emph {et~al.}(1986)\citenamefont
  {Rumelhart}, \citenamefont {Hinton},\ and\ \citenamefont
  {Williams}}]{Rumelhart:1986we}%
  \BibitemOpen
  \bibfield  {author} {\bibinfo {author} {\bibfnamefont {David~E.}\
  \bibnamefont {Rumelhart}}, \bibinfo {author} {\bibfnamefont {Geoffrey~E.}\
  \bibnamefont {Hinton}}, \ and\ \bibinfo {author} {\bibfnamefont {Ronald~J.}\
  \bibnamefont {Williams}},\ }\bibfield  {title} {\enquote {\bibinfo {title}
  {{Learning Representations by Back-propagating Errors}},}\ }\href {\doibase
  10.1038/323533a0} {\bibfield  {journal} {\bibinfo  {journal} {Nature}\
  }\textbf {\bibinfo {volume} {323}},\ \bibinfo {pages} {533--536} (\bibinfo
  {year} {1986})}\BibitemShut {NoStop}%
\bibitem [{\citenamefont {Buckley}\ \emph {et~al.}(2011)\citenamefont {Buckley}
  \emph {et~al.}}]{Buckley:2011ms}%
  \BibitemOpen
  \bibfield  {author} {\bibinfo {author} {\bibfnamefont {Andy}\ \bibnamefont
  {Buckley}} \emph {et~al.},\ }\bibfield  {title} {\enquote {\bibinfo {title}
  {{General-purpose event generators for LHC physics}},}\ }\href {\doibase
  10.1016/j.physrep.2011.03.005} {\bibfield  {journal} {\bibinfo  {journal}
  {Phys. Rept.}\ }\textbf {\bibinfo {volume} {504}},\ \bibinfo {pages}
  {145--233} (\bibinfo {year} {2011})},\ \Eprint
  {http://arxiv.org/abs/1101.2599} {arXiv:1101.2599 [hep-ph]} \BibitemShut
  {NoStop}%
\bibitem [{\citenamefont {H\"oche}(2015)}]{Hoche:2014rga}%
  \BibitemOpen
  \bibfield  {author} {\bibinfo {author} {\bibfnamefont {Stefan}\ \bibnamefont
  {H\"oche}},\ }\bibfield  {title} {\enquote {\bibinfo {title} {{Introduction
  to parton-shower event generators}},}\ }in\ \href {\doibase
  10.1142/9789814678766_0005} {\emph {\bibinfo {booktitle} {{Theoretical
  Advanced Study Institute in Elementary Particle Physics}: {Journeys Through
  the Precision Frontier: Amplitudes for Colliders}}}}\ (\bibinfo {year}
  {2015})\ pp.\ \bibinfo {pages} {235--295},\ \Eprint
  {http://arxiv.org/abs/1411.4085} {arXiv:1411.4085 [hep-ph]} \BibitemShut
  {NoStop}%
\bibitem [{\citenamefont {Andersson}\ \emph {et~al.}(1996)\citenamefont
  {Andersson}, \citenamefont {Gustafson},\ and\ \citenamefont
  {Samuelsson}}]{ANDERSSON1996217}%
  \BibitemOpen
  \bibfield  {author} {\bibinfo {author} {\bibfnamefont {B.}~\bibnamefont
  {Andersson}}, \bibinfo {author} {\bibfnamefont {G.}~\bibnamefont
  {Gustafson}}, \ and\ \bibinfo {author} {\bibfnamefont {J.}~\bibnamefont
  {Samuelsson}},\ }\bibfield  {title} {\enquote {\bibinfo {title} {Discrete
  qcd, a new approximation for qcd cascades},}\ }\href {\doibase
  https://doi.org/10.1016/0550-3213(96)00022-3} {\bibfield  {journal} {\bibinfo
   {journal} {Nuclear Physics B}\ }\textbf {\bibinfo {volume} {463}},\ \bibinfo
  {pages} {217--237} (\bibinfo {year} {1996})}\BibitemShut {NoStop}%
\bibitem [{\citenamefont {Gustafson}\ \emph {et~al.}(2022)\citenamefont
  {Gustafson}, \citenamefont {Prestel}, \citenamefont {Spannowsky},\ and\
  \citenamefont {Williams}}]{Gustafson:2022xwt}%
  \BibitemOpen
  \bibfield  {author} {\bibinfo {author} {\bibfnamefont {G\"osta}\ \bibnamefont
  {Gustafson}}, \bibinfo {author} {\bibfnamefont {Stefan}\ \bibnamefont
  {Prestel}}, \bibinfo {author} {\bibfnamefont {Michael}\ \bibnamefont
  {Spannowsky}}, \ and\ \bibinfo {author} {\bibfnamefont {Simon}\ \bibnamefont
  {Williams}},\ }\bibfield  {title} {\enquote {\bibinfo {title} {{Collider
  Events on a Quantum Computer}},}\ }\href@noop {} {\  (\bibinfo {year}
  {2022})},\ \Eprint {http://arxiv.org/abs/2207.10694} {arXiv:2207.10694
  [hep-ph]} \BibitemShut {NoStop}%
\bibitem [{\citenamefont {Farhi}(1977)}]{Farhi:1977sg}%
  \BibitemOpen
  \bibfield  {author} {\bibinfo {author} {\bibfnamefont {Edward}\ \bibnamefont
  {Farhi}},\ }\bibfield  {title} {\enquote {\bibinfo {title} {{A QCD Test for
  Jets}},}\ }\href {\doibase 10.1103/PhysRevLett.39.1587} {\bibfield  {journal}
  {\bibinfo  {journal} {Phys. Rev. Lett.}\ }\textbf {\bibinfo {volume} {39}},\
  \bibinfo {pages} {1587--1588} (\bibinfo {year} {1977})}\BibitemShut {NoStop}%
\bibitem [{\citenamefont {Andersson}\ \emph {et~al.}(1983)\citenamefont
  {Andersson}, \citenamefont {Gustafson}, \citenamefont {Ingelman},\ and\
  \citenamefont {Sjostrand}}]{Andersson:1983ia}%
  \BibitemOpen
  \bibfield  {author} {\bibinfo {author} {\bibfnamefont {Bo}~\bibnamefont
  {Andersson}}, \bibinfo {author} {\bibfnamefont {G.}~\bibnamefont
  {Gustafson}}, \bibinfo {author} {\bibfnamefont {G.}~\bibnamefont {Ingelman}},
  \ and\ \bibinfo {author} {\bibfnamefont {T.}~\bibnamefont {Sjostrand}},\
  }\bibfield  {title} {\enquote {\bibinfo {title} {{Parton Fragmentation and
  String Dynamics}},}\ }\href {\doibase 10.1016/0370-1573(83)90080-7}
  {\bibfield  {journal} {\bibinfo  {journal} {Phys. Rept.}\ }\textbf {\bibinfo
  {volume} {97}},\ \bibinfo {pages} {31--145} (\bibinfo {year}
  {1983})}\BibitemShut {NoStop}%
\bibitem [{\citenamefont {Bierlich}\ \emph {et~al.}(2022)\citenamefont
  {Bierlich} \emph {et~al.}}]{Bierlich:2022pfr}%
  \BibitemOpen
  \bibfield  {author} {\bibinfo {author} {\bibfnamefont {Christian}\
  \bibnamefont {Bierlich}} \emph {et~al.},\ }\bibfield  {title} {\enquote
  {\bibinfo {title} {{A comprehensive guide to the physics and usage of PYTHIA
  8.3}},}\ }\href@noop {} {\  (\bibinfo {year} {2022})},\ \Eprint
  {http://arxiv.org/abs/2203.11601} {arXiv:2203.11601 [hep-ph]} \BibitemShut
  {NoStop}%
\bibitem [{\citenamefont {Heister}\ \emph {et~al.}(2004)\citenamefont {Heister}
  \emph {et~al.}}]{ALEPH:2003obs}%
  \BibitemOpen
  \bibfield  {author} {\bibinfo {author} {\bibfnamefont {A.}~\bibnamefont
  {Heister}} \emph {et~al.} (\bibinfo {collaboration} {ALEPH}),\ }\bibfield
  {title} {\enquote {\bibinfo {title} {{Studies of QCD at $e^+ e^-$
  centre-of-mass energies between 91 GeV and 209 GeV}},}\ }\href {\doibase
  10.1140/epjc/s2004-01891-4} {\bibfield  {journal} {\bibinfo  {journal} {Eur.
  Phys. J. C}\ }\textbf {\bibinfo {volume} {35}},\ \bibinfo {pages} {457--486}
  (\bibinfo {year} {2004})}\BibitemShut {NoStop}%
\bibitem [{\citenamefont {De~Castro}\ and\ \citenamefont
  {Dorigo}(2019)}]{deCastro:2018mgh}%
  \BibitemOpen
  \bibfield  {author} {\bibinfo {author} {\bibfnamefont {Pablo}\ \bibnamefont
  {De~Castro}}\ and\ \bibinfo {author} {\bibfnamefont {Tommaso}\ \bibnamefont
  {Dorigo}},\ }\bibfield  {title} {\enquote {\bibinfo {title} {{INFERNO:
  Inference-Aware Neural Optimisation}},}\ }\href {\doibase
  10.1016/j.cpc.2019.06.007} {\bibfield  {journal} {\bibinfo  {journal}
  {Comput. Phys. Commun.}\ }\textbf {\bibinfo {volume} {244}},\ \bibinfo
  {pages} {170--179} (\bibinfo {year} {2019})},\ \Eprint
  {http://arxiv.org/abs/1806.04743} {arXiv:1806.04743 [stat.ML]} \BibitemShut
  {NoStop}%
\bibitem [{\citenamefont {Simpson}\ and\ \citenamefont
  {Heinrich}(2022)}]{Simpson:2022suz}%
  \BibitemOpen
  \bibfield  {author} {\bibinfo {author} {\bibfnamefont {Nathan}\ \bibnamefont
  {Simpson}}\ and\ \bibinfo {author} {\bibfnamefont {Lukas}\ \bibnamefont
  {Heinrich}},\ }\bibfield  {title} {\enquote {\bibinfo {title} {{neos:
  End-to-End-Optimised Summary Statistics for High Energy Physics}},}\ }in\
  \href {\doibase 10.48550/arXiv.2203.05570} {\emph {\bibinfo {booktitle}
  {{20th International Workshop on Advanced Computing and Analysis Techniques
  in Physics Research}: {AI Decoded - Towards Sustainable, Diverse, Performant
  and Effective Scientific Computing}}}}\ (\bibinfo {year} {2022})\ \Eprint
  {http://arxiv.org/abs/2203.05570} {arXiv:2203.05570 [physics.data-an]}
  \BibitemShut {NoStop}%
\bibitem [{\citenamefont {Gustafson}(1993)}]{Gustafson:1992uh}%
  \BibitemOpen
  \bibfield  {author} {\bibinfo {author} {\bibfnamefont {Gosta}\ \bibnamefont
  {Gustafson}},\ }\bibfield  {title} {\enquote {\bibinfo {title} {{Multiplicity
  distributions in QCD cascades}},}\ }\href {\doibase
  10.1016/0550-3213(93)90203-2} {\bibfield  {journal} {\bibinfo  {journal}
  {Nucl. Phys. B}\ }\textbf {\bibinfo {volume} {392}},\ \bibinfo {pages}
  {251--280} (\bibinfo {year} {1993})}\BibitemShut {NoStop}%
\bibitem [{\citenamefont {Campbell}\ \emph {et~al.}(2022)\citenamefont
  {Campbell} \emph {et~al.}}]{Campbell:2022qmc}%
  \BibitemOpen
  \bibfield  {author} {\bibinfo {author} {\bibfnamefont {J.~M.}\ \bibnamefont
  {Campbell}} \emph {et~al.},\ }\bibfield  {title} {\enquote {\bibinfo {title}
  {{Event Generators for High-Energy Physics Experiments}},}\ }in\ \href@noop
  {} {\emph {\bibinfo {booktitle} {{2022 Snowmass Summer Study}}}}\ (\bibinfo
  {year} {2022})\ \Eprint {http://arxiv.org/abs/2203.11110} {arXiv:2203.11110
  [hep-ph]} \BibitemShut {NoStop}%
\end{thebibliography}%

\end{document}